\documentclass{article}

\usepackage[preprint]{neurips_2020}

\usepackage[utf8]{inputenc} % allow utf-8 input
\usepackage[T1]{fontenc}    % use 8-bit T1 fonts
\usepackage{hyperref}       % hyperlinks
\usepackage{url}            % simple URL typesetting
\usepackage{booktabs}       % professional-quality tables
\usepackage{amsfonts}       % blackboard math symbols
\usepackage{nicefrac}       % compact symbols for 1/2, etc.
\usepackage{microtype}      % microtypography

\usepackage{amssymb,amsmath,amsthm,enumerate,threeparttable,bm,subfigure,graphicx}
\usepackage{algorithm, algorithmic}
\usepackage{booktabs}
\usepackage{multirow}
\usepackage{caption}
\usepackage{wrapfig}
\usepackage{bm}
\usepackage{threeparttable}
\usepackage{mathrsfs}
\usepackage{lipsum}
\usepackage{enumitem}
\usepackage{multirow, subfigure}
\usepackage{appendix}
\usepackage{hyperref}       % hyperlinks
\usepackage{color}

\newtheorem{definition}{Definition}

\usepackage{color}

\long\def\comment#1{}

\def\ie{$i.e.$}

\title{Open-sourced Dataset Protection via Backdoor Watermarking}

\author{%
     Yiming Li$^{1}$, Ziqi Zhang$^{1}$, Jiawang Bai$^{1}$, Baoyuan Wu$^{2,3}$, Yong Jiang$^{1}$, Shu-Tao Xia$^{1,}$\thanks{Correspondence to: Shu-Tao Xia (\texttt{xiast@sz.tsinghua.edu.cn})}\\
$^1$Tsinghua Shenzhen International Graduate School, Tsinghua University, China\\
$^2$ School of Data Science, The Chinese University of Hong Kong, Shenzhen, China \\
$^3$ Secure Computing Lab of Big Data, Shenzhen Research Institute of Big Data, China \\
\texttt{li-ym18@mails.tsinghua.edu.cn}; \texttt{xiast@sz.tsinghua.edu.cn}\\
}

\begin{document}

\maketitle

\vspace{-1em}
\begin{abstract}
\vspace{-0.4em}
  The rapid development of deep learning has benefited from the release of some high-quality open-sourced datasets ($e.g.$, ImageNet), which allows researchers to easily verify the effectiveness of their algorithms. Almost all existing open-sourced datasets require that they can only be adopted for academic or educational purposes rather than commercial purposes, whereas there is still no good way to protect them. In this paper, we propose a \emph{backdoor embedding based dataset watermarking} method to protect an open-sourced image-classification dataset by verifying whether it is used for training a third-party model. Specifically, the proposed method contains two main processes, including \emph{dataset watermarking} and \emph{dataset verification}. We adopt classical poisoning-based backdoor attacks ($e.g.$, BadNets) for dataset watermarking, \ie, generating some poisoned samples by adding a certain trigger ($e.g.$, a local patch) onto some benign samples, labeled with a pre-defined target class. Based on the proposed backdoor-based watermarking, we use a hypothesis test guided method for dataset verification based on the posterior probability generated by the suspicious third-party model of the benign samples and their correspondingly watermarked samples ($i.e.$, images with trigger) on the target class. Experiments on some benchmark datasets are conducted, which verify the effectiveness of the proposed method. %The code for reproducing main results is available at \url{https://github.com/THUYimingLi/Open-sourced_Dataset_Protection}.
  
\end{abstract}

\vspace{-1.2em}
\section{Introduction}
\vspace{-0.6em}

Recently, deep neural networks (DNNs) have been widely and successfully adopted in many fields \cite{Zhu2020TheRO, zhang2019hibert, ren2019fastspeech, liu2020transferring, C_H@TPAMI_2020, Wu_2019_ICCV} for their extraordinary performance. Datasets, especially high-quality open-sourced datasets ($e.g.$, ImageNet \cite{deng2009imagenet}, CIFAR \cite{cifar}), are the key factors of the prosperity of DNNs. Those datasets allow researchers to easily verify the effectiveness of their algorithms or models, which in turn accelerates the development of deep learning. Since the collection of datasets is time-consuming and even expensive, all datasets deserve to be protected. 
Some techniques of dataset protection have been proposed, such as anonymization \cite{k-anon@2002, ZK@2019, YS@2019}, encryption \cite{HUA2019403, CHEN2019420, LI2019113}, and watermarking \cite{arsalan2017protection, chai_2019, D_Hu_2019}. 
Their purposes are precluding unauthorized users to access the dataset. However, they are not suitable to protect open-sourced datasets. Because, many open-sourced datasets are open access to everyone, with the only requirement that they can only be adopted for academic or educational rather than commercial purposes. 
Thus, the main challenge for protecting open-sourced datasets is verifying whether it has been adopted for training a third-party model without given any training details.
To the best of our knowledge, there is no prior work to tackle this challenge.

In this paper, we propose a novel method for protecting open-sourced datasets, dubbed \emph{backdoor embedding based dataset watermarking (BEDW)}. It consists of two main processes, including \emph{dataset watermarking} and \emph{dataset verification}. Specifically, we adopt poisoning-based backdoor attacks \cite{chen2017targeted,gu2019badnets} for dataset watermarking, inspired by the fact that those methods can create an attacker-specified hidden backdoor while maintaining the prediction accuracy on benign samples, when the model is trained on the watermarked dataset with the standard training process. %Those backdoor attacks are stealthy and effective therefore can be used by open-sourced dataset protectors. 
Moreover, based on the proposed backdoor-based dataset watermarking, protectors can verify whether the model was trained on their dataset by examining the existence of the specific backdoor. Specifically, since the hidden backdoor can be regarded as the relation between the backdoor trigger and the target label, the protector only needs to verify whether the posterior probability on the target class of watermarked samples is significantly higher than that of benign samples. To this end, we adopt a pairwise hypothesis test guided method to verify it.

The main contributions of this work are three-fold. 
{\bf 1)} We propose to protect the open-sourced datasets by verifying whether the dataset has been adopted to train a third-party model. 
{\bf 2)} We present a novel dataset watermarking method for protecting open-sourced datasets, based on the properties of poisoning-based backdoor attacks and pair-wise hypothesis tests. 
{\bf 3)} Extensive experiments verify the effectiveness of the proposed method. 
  
\comment{
\begin{itemize}[leftmargin=2.5em]
    \item We explore how to protect the open-sourced dataset. Specifically, we formulate this problem as determining whether a given model is trained on the protected dataset based on its predictions. 
    \item We proposed a new dataset protection method inspired by the properties of poisoning-based attacks and pair-wise hypothesis tests.
    \item Extensive experiments verify the effectiveness of the proposed method. 
\end{itemize}
}

\vspace{-0.8em}
\section{The Proposed Method}
\vspace{-0.6em}
In this section, we explore how to protect a dataset, especially an open-sourced dataset, based on the backdoor watermarking. Specifically, the proposed method 
 contains two main processes, including \emph{dataset watermarking} and \emph{dataset verification}. They will be discussed in detail in this section.

\vspace{-0.5em}
\subsection{Dataset Watermarking based on Poisoning-based Backdoor Attack}\label{sec:watermark}
\vspace{-0.4em}

Intuitively, the proposed watermarking method needs to satisfy three main properties, including the \emph{harmlessness}, \emph{effectiveness}, and \emph{stealthiness}, as follows:

\vspace{0.3em}
\begin{definition}[Three Necessary Properties for the Dataset Watermarking]
\end{definition}
\vspace{-0.8em}
\begin{itemize}[leftmargin=2em]
    \item \textbf{Harmlessness}: \emph{The watermarking should not hinder its normal usage, $i.e.,$ the performance of models trained on the watermarked dataset should be on par with those on the benign dataset.}
    \item \textbf{Effectiveness}: \emph{The model trained with the watermarked dataset will be `marked' with a certain imprint, which can be used to verify whether the dataset has been stolen for training.}
    \item \textbf{Stealthiness}: \emph{The watermarking should not attract the attention of the dataset stealer. }
\end{itemize}

\vspace{-0.2em}
Adopting poisoning-based backdoor attacks \cite{gu2019badnets,bagdasaryan2020backdoor,li2020rethinking}, especially invisible backdoor attacks \cite{chen2017targeted,liu2020reflection,li2020invisible}, would be a good option to fulfill all those requirements. Specifically, in the poisoning-based backdoor attack, a small number of training samples are modified by adding the trigger ($e.g.,$ the local patch). These modified samples with attacker-specified target label, together with benign training samples, are formed as the watermarked training dataset. Accordingly, the prediction of models trained with the watermarked dataset under standard training process will be maliciously changed if the hidden backdoor is activated by the attacker-defined trigger, while they perform well on benign samples \cite{liu2020survey,li2020backdoor}. As such, the backdoor-based watermarking is harmless and stealth, and the protector-specified `trigger--target label' relation can be used to verify whether the dataset is stolen for commercial purposes. In this paper, we adopt the \emph{BadNets} \cite{gu2019badnets} and the invisible attack with blended strategy \cite{chen2017targeted} (dubbed \emph{Blended Attack}) to watermark the dataset. More exploration about watermarking with other attacks will be discussed in our future work.  

\textbf{The Generation of the Watermarked Dataset. }
Suppose $\mathcal{D}_{train} = \{ (\bm{x}_i, y_i) \}_{i=1}^N$ indicates the benign ($i.e.$, unprotected) training set where $\bm{x}_i \in \{0,\cdots, 255\}^{C\times W \times H}$ and $y_i \in \{1,\cdots, K\}$. Let $y_t \in \{1,\cdots, K\}$ and $\bm{t} \in \{0,\cdots, 255\}^{C\times W \times H}$ be the target label and the protector-specified trigger, respectively. Note that $\bm{t}$ itself should not semantically correlate to $y_{target}$. Watermarking a benign image $\bm{x}_{benign}$ with BadNets or Blended Attack can be denoted as follows:

\begin{equation}
\bm{x}_{watermarked} = w(\bm{x}_{benign}; \bm{t}) = (\bm{1}-\bm{\lambda}) \otimes \bm{x}_{benign} + \bm{\lambda} \otimes \bm{t}, 
\end{equation}
where $\bm{\lambda} \in [0,1]^{C \times W \times H}$ is a visibility-related hyper-parameter and $\otimes$ indicates the element-wise product. The smaller the $\bm{\lambda}$, the more invisible the trigger and the more stealthy the watermarking.

The watermarked dataset $\mathcal{D}_{watermarked}$ are the union of remaining benign training samples and a small number of watermarked training images with target label, $i.e.$, 
\begin{equation}
   \mathcal{D}_{watermarked} = \mathcal{D}_{benign} \cup \mathcal{D}_{modified},
\end{equation}
where $\mathcal{D}_{modified} = \left\{(\bm{x}', y_{target})| \bm{x}' = w(\bm{x}; \bm{t}), (\bm{x},y) \in \mathcal{D}_{train} \backslash \mathcal{D}_{benign} \right\}$,  $\mathcal{D}_{benign} \subset \mathcal{D}_{train}$, $\gamma =\frac{|\mathcal{D}_{modified}|}{|\mathcal{D}_{train}|}$ is the watermarking rate. The smaller the $\gamma$, the more stealthy the watermark.

\begin{table}[ht]
\centering
\small
\vspace{-1.5em}
\caption{The effectiveness of dataset watermarking (with $5\%$ watermarking rate) measured by BA and WSR on CIFAR. `BA' and `WSR' indicate the benign accuracy ($i.e.$, accuracy on benign testing set) and watermark success rate (accuracy on 100\% watermarked testing set), respectively. }
\vspace{0.15em}
\scalebox{0.9}{
\begin{tabular}{c|cc|cc|cc|cc|cc}
\hline
Watermark $\rightarrow$ & \multicolumn{2}{c|}{No Watermark}  & \multicolumn{4}{c|}{BadNets}                                        & \multicolumn{4}{c}{Blended Attack ($\bm{\lambda} \in \{0, 0.2\}^{C\times W \times H}$)}                                    \\ \hline
Trigger $\rightarrow$ & \multicolumn{2}{c|}{No Trigger} & \multicolumn{2}{c|}{Black Line} & \multicolumn{2}{c|}{White Square} & \multicolumn{2}{c|}{Black Line} & \multicolumn{2}{c}{White Square}                     \\ \hline
Model $\downarrow$  & Train ACC       & Test ACC      & BA             & WSR            & BA              & WSR             & BA             & WSR            & BA                        & WSR                       \\ \hline
ResNet  & 100        & 92.21         & 91.63          & 99.51          & 91.86           & 97.47           & 91.85          & 94.01          & 91.81                     & 84.99                     \\ \hline
VGG     & 99.98           & 92.30         & 91.67          & 99.25          & 91.97           & 96.78           & 91.38          & 92.86         & 90.87  & 77.68 \\ \hline
\end{tabular}
}
\vspace{-1.7em}
\label{tab:WSR_CIFAR}
\end{table}

\vspace{-0.4em}
\subsection{Dataset Verification with Pairwise Hypothesis Test}
\vspace{-0.4em}
\label{sec:veri}
Given a trained third-party classifier, protectors can verify whether the model was trained with their (watermarked) dataset through examining the existence of the specific backdoor. Specifically, since the hidden backdoor can be regarded as the relation between the backdoor trigger and the target label, the protector only needs to verify whether the posterior probability on the target class of watermarked testing samples is significantly higher than that of benign testing samples.

Given a classifier $C(\cdot; \bm{\theta})$, let $f(\bm{x};\bm{\theta})$ indicates the posterior probability of $\bm{x}$ predicted by the classifier. Suppose variable $\bm{X}$ indicates the benign image from a class with a label different from $y_{target}$, variable $\bm{X}' = w(\bm{X}; \bm{t})$ denotes the watermarked version of $\bm{X}$. Let variable $p=f(\bm{X};\bm{\theta})_{y_{target}}$ and $q=f(\bm{X}';\bm{\theta})_{y_{target}}$ indicate the posterior probability generated by classifier $C(\cdot; \bm{\theta})$ on the target label $y_{target}$ of $\bm{X}$ and $\bm{X}'$, respectively. 

We propose a hypothesis test based dataset verification to verify the existence of backdoor for determining whether the dataset is stolen, as follows:

\emph{Given the null hypothesis $H_0: p + \alpha > q$ ($H_1: p + \alpha \leq q$) where the hyper-parameter $\alpha \in [0,1]$, we claim that the model is trained on the watermarked dataset (with $\alpha$-certainty) iff $H_0$ is rejected.} 

%Note that $\alpha \in [0,1]$ is a hyper-parameter. The higher the $\alpha$, the greater the authenticity of the identified backdoor, while the easier it is to miss the backdoor.

To conduct the dataset verification, suppose that the protector generates a dataset $\{\bm{x}_i\}_{i=1}^M$ containing $M$ $i.i.d.$ observations of $\bm{X}$ and let $\{\bm{x}_i'\}_{i=1}^M$ indicates their correspondingly watermarked version, $i.e.$, $\forall i, \bm{x}_i' = w(\bm{x}_{i}; \bm{t})$. The protector can examine the null hypothesis $H_0$ through a pairwise hypothesis test. Since variable $q-p$ and the effect of the backdoor trigger are closely related, $q-p$ can be approximately regarded as following the Gaussian distribution \cite{hogg2005introduction}. Accordingly, we adopt the \emph{pairwise T-test} \cite{hogg2005introduction} for the dataset verification in this paper. One may also adopt \emph{pairwise Wilcoxon-test} \cite{wilcoxon1992individual} when the observations are not $i.i.d.$, which will be discussed in our future work.

\vspace{-1em}
\section{Experiments}
\vspace{-0.8em}

\subsection{Main Results}\label{sec:main_results}
\vspace{-0.2em}

\textbf{Evaluation Metric. } We adopt benign accuracy (BA) and watermark success rate (WSR) to verify the effectiveness of the dataset watermarking. BA is defined as the model accuracy on the benign testing set, while the WSR indicates the model accuracy on the 100\% watermarked testing set. When examining the effectiveness of the dataset verification, we verify the existence of hidden backdoor in models trained on the watermarked dataset multiple times with the method proposed in Section \ref{sec:veri}. We calculate the ratio of successful detections (RSD) as its effectiveness measurement. The higher the BA, WSR, and RSD, the better the performance.

\begin{wrapfigure}{r}{0cm}
\centering
\small
\includegraphics[width=0.4\textwidth]{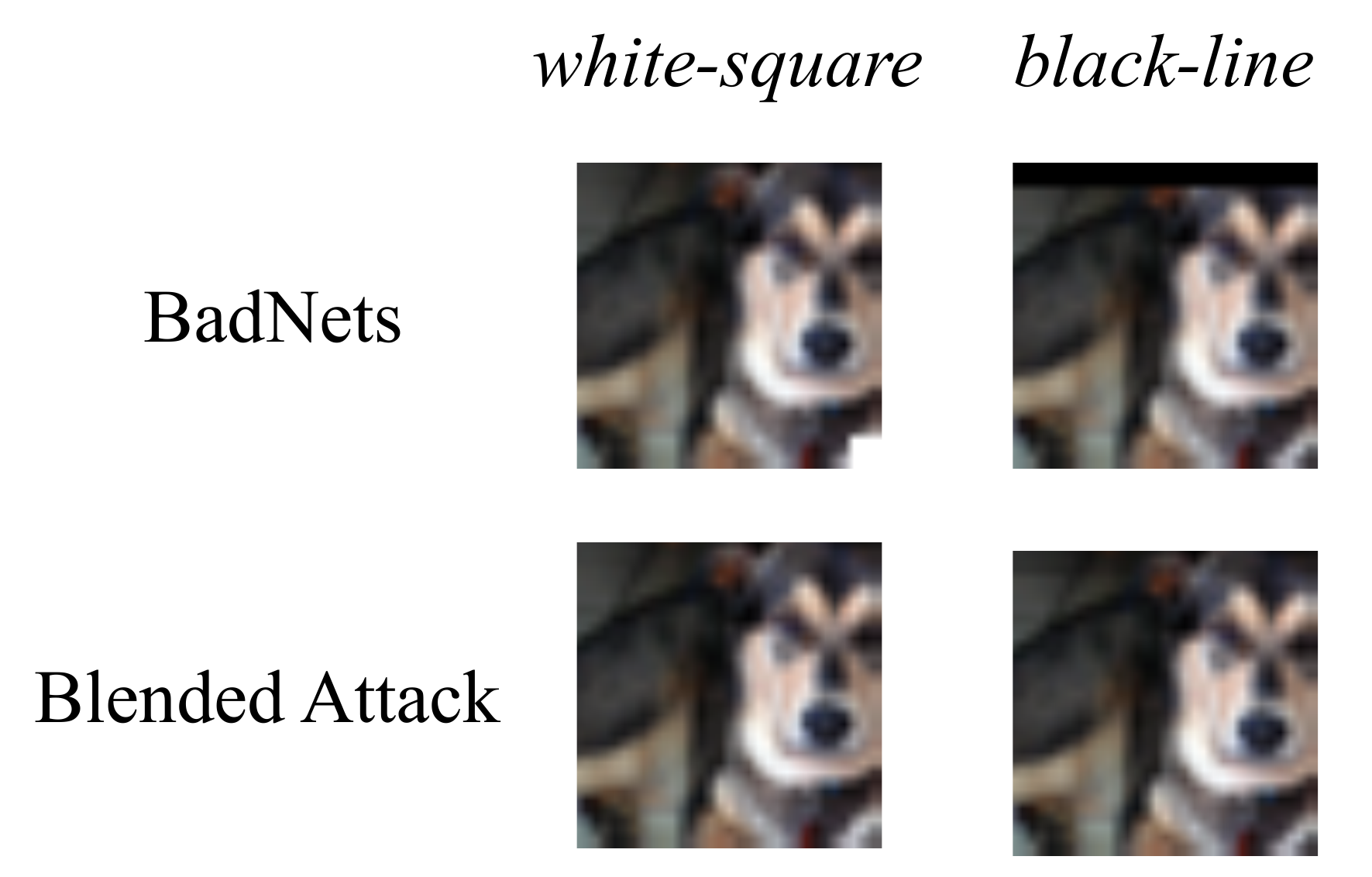}
\caption{Watermarked images generated by BadNets and Blended Attack with different trigger patterns.}
\label{fig:watermarked}
\vspace{-1em}
\end{wrapfigure}

\textbf{Settings. } We use VGG-19 \cite{simonyan2014very} and ResNet-18 \cite{he2016deep} as the model structure and conduct experiments on CIFAR-10 \cite{cifar} and GTSRB \cite{gtsrb} dataset. For the dataset watermarking, we examine two types of triggers, including $3\times3$ white square and black line with 3-pixels width. We set watermarking rate $\gamma=0.05$ and $\bm{\lambda} \in \{0, 0.2\}^{C \times W \times H}$ in the blended attack.
The values of $\bm{\lambda}$ entries corresponding to the pixels located in trigger area are 0.2, while others are 0. Besides, the target label $y_{target}$ is set to `1' on both CIFAR-10 and GTSRB dataset. For the dataset verification, we set $\alpha=0.5$ and randomly sample $M=100$ samples with the same class different from $y_{target}$ and repeat the experiments 100 times to calculate the RSD. The pairwise T-test is performed at a significance level of 0.05. Some watermarked images are shown in Figure \ref{fig:watermarked}.

\begin{table}[ht]
\centering
\small
\caption{The effectiveness of dataset watermarking (with $5\%$ watermarking rate) measured by BA and WSR (\%) on GTSRB. `BA' and `WSR' indicate the benign accuracy ($i.e.$, accuracy on benign testing set) and watermark success rate (accuracy on 100\% watermarked testing set), respectively. }
\vspace{0.15em}
\scalebox{0.9}{
\begin{tabular}{c|cc|cc|cc|cc|cc}
\hline
Watermark $\rightarrow$    & \multicolumn{2}{c|}{No Watermark}  & \multicolumn{4}{c|}{BadNets}                                                              & \multicolumn{4}{c}{Blended Attack ($\bm{\lambda} \in \{0, 0.2\}^{C \times W \times H}$)}                                            \\ \hline
Trigger $\rightarrow$   & \multicolumn{2}{c|}{No Trigger} & \multicolumn{2}{c|}{Black Line} & \multicolumn{2}{c|}{White Square}                       & \multicolumn{2}{c|}{Black Line} & \multicolumn{2}{c}{White Square}                             \\ \hline
Model $\downarrow$    & Train ACC       & Test ACC      & BA             & WSR            & BA                         & WSR                        & BA             & WSR            & BA                            & WSR                           \\ \hline
ResNet & 100             & 98.35         & 97.76          & 94.70          & 97.82 & 79.37 & 97.83          & 94.14          & 97.61 & 86.94 \\ \hline
VGG       & 100             & 97.06         & 96.97          & 98.96         & 96.83 & 74.01 & 96.37          & 91.96          & 97.04                         & 84.58                         \\ \hline
\end{tabular}
}
\vspace{-1.25em}
\label{tab:WSR_GTSRB}
\end{table}

\begin{table}[!ht]
\centering
\small
\caption{The effectiveness of the dataset verification measured by the ratio of successful detections (RSD) (\%) with certainty $\alpha=0.5$ towards models trained on different watermarked datasets.}
\vspace{0.15em}
\begin{tabular}{c|cc|cc|cc|cc}
\hline
Dataset $\rightarrow$       & \multicolumn{4}{c|}{CIFAR}                                            & \multicolumn{4}{c}{GTSRB}                                            \\ \hline

Watermark $\rightarrow$        & \multicolumn{2}{c|}{BadNets}    & \multicolumn{2}{c|}{Blended Attack} & \multicolumn{2}{c|}{BadNets}    & \multicolumn{2}{c}{Blended Attack} \\ \hline
Model $\downarrow$,\ Trigger $\rightarrow$ & Line          & Square          & Line            & Square            & Line          & Square          & Line            & Square            \\ \hline
ResNet         &  100             &  100               & 100                &   100                &  100     & 100                &  100               &   100                \\ \hline
VGG            &   100            &   100              &  100               &  100                 &  100            &  100               &  100               &  100                 \\ \hline
\end{tabular}
\label{tab:RSD}
\vspace{-0.8em}
\end{table}

\begin{figure}[!ht]
\begin{minipage}[b]{0.49\linewidth}
    \centering
    \includegraphics[width=0.9\textwidth]{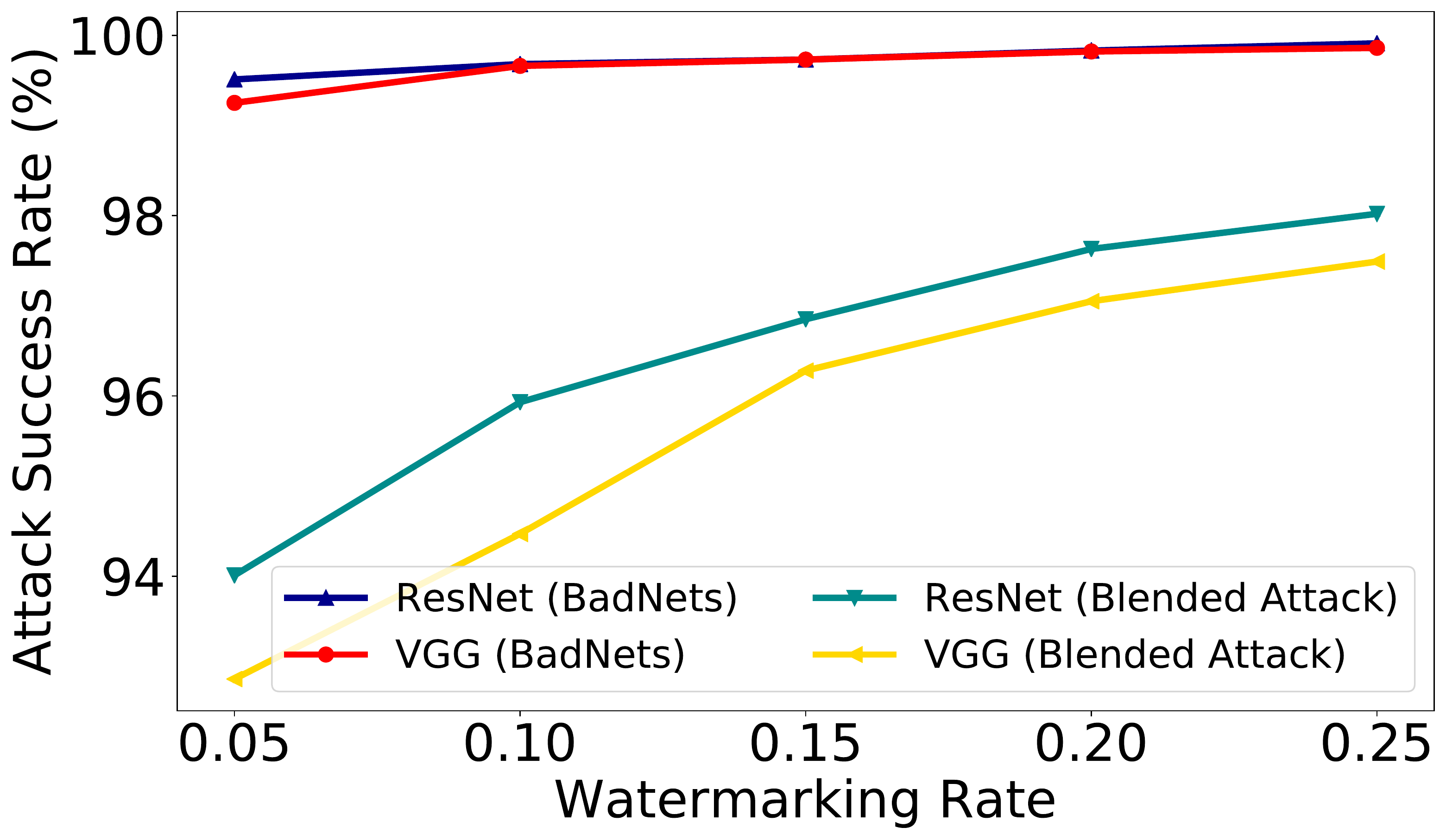}
    \caption{The ASR $w.r.t.$ different watermarking rate $\gamma$ with line-type trigger on CIFAR-10 dataset.}
    \label{fig:wr}
\label{fig_position}
\end{minipage}\quad
\begin{minipage}[b]{0.47\linewidth}
    \centering
    \includegraphics[width=0.9\textwidth]{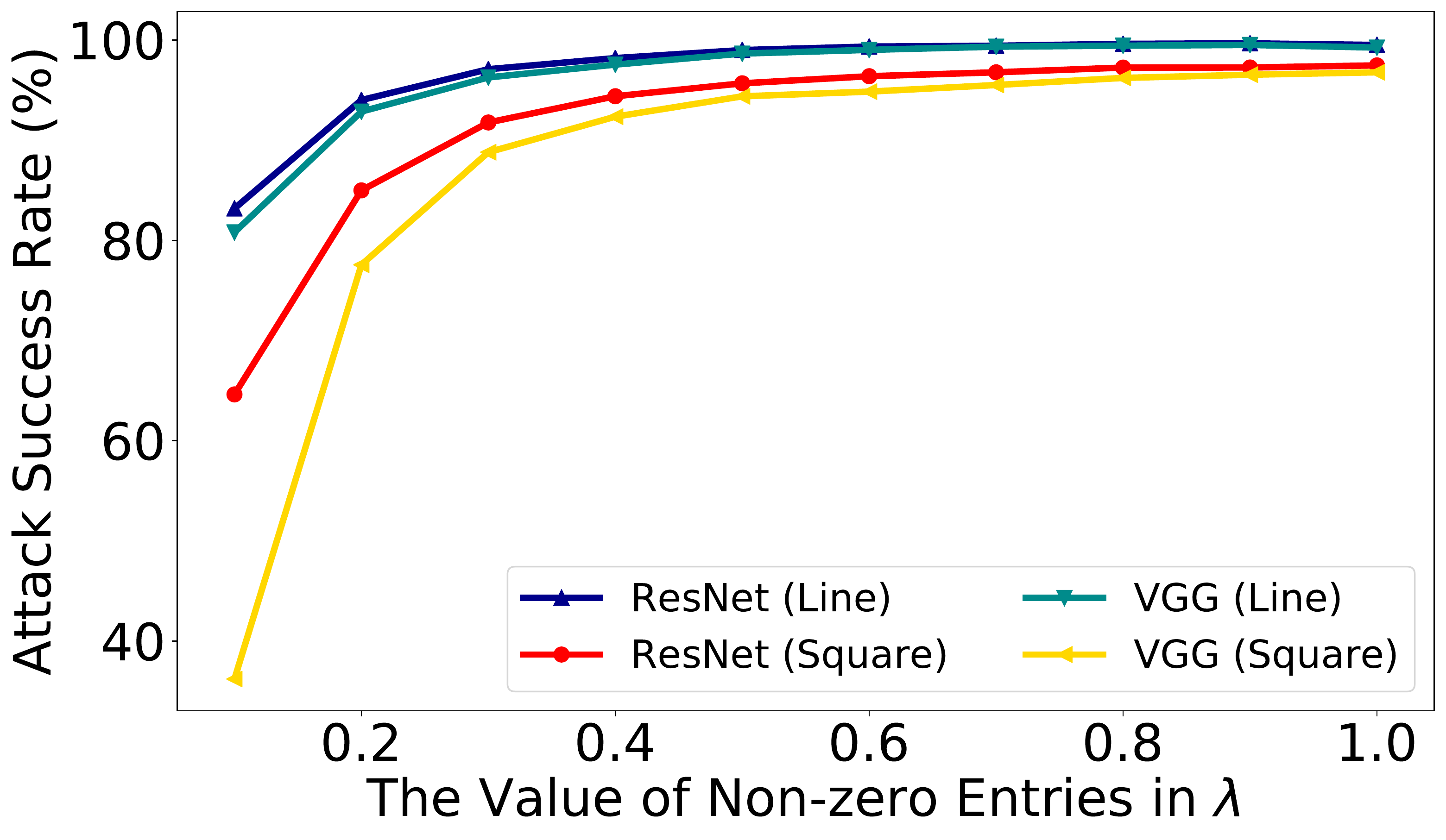}
    \caption{The ASR $w.r.t.$ different visibility-related parameter $\bm{\lambda}$ on CIFAR-10 dataset.}
     \label{fig:lambda}
\end{minipage}
\vspace{-1.2em}
\end{figure}

\textbf{Results. } As shown in Table \ref{tab:WSR_CIFAR}-\ref{tab:WSR_GTSRB}, compared with the performance of those trained on benign training sets, the decreases of benign accuracy of models trained on watermarked training sets are all less than 2\%. Besides, the WSR of models under most cases are greater than $90\%$ with only a 5\% watermarking rate. Those results verify the harmlessness, effectiveness, and stealthiness of the proposed dataset watermarking. Moreover, Table \ref{tab:RSD} also shows the remarkable performance of the proposed dataset verification. The proposed verification method can accurately identify the backdoor in DNNs trained with watermarked datasets, even if the WSR does not reach 100\%.

\vspace{-0.4em}
\subsection{The Ablation Study}
\vspace{-0.3em}
In this section, we discuss the effect of two hyper-parameters ($i.e.$, the watermarking rate $\gamma$, and the visibility-related hyper-parameter $\bm{\lambda}$) toward the WSR in the proposed method. Except for the studied hyper-parameters, other settings are the same as those used in Section \ref{sec:main_results}.

As shown in Figure \ref{fig:wr}, the WSR increases along with an increase of $\gamma$. However, its increase will also decrease the stealthiness of the dataset watermarking. The $\gamma$ should be specified based on the needs of protectors. 
Besides, Figure \ref{fig:lambda} shows that the WSR also increases with an increase of the value of non-zero entries in $\bm{\lambda}$. This phenomenon may be due to the reason that it is easier for DNNs to learn `robust features' \cite{ilyas2019adversarial}. We will further explore it in our future work.

\vspace{-0.8em}
\section{Conclusion}
\vspace{-0.6em}

In this paper, we have studied how to protect open-sourced datasets by ensuring that they are not used for inappropriate purposes. We provided a novel perspective that this task could be formulated as verifying whether a given model is trained on the protected dataset. 
To this end, we proposed a backdoor embedding based dataset watermarking method, consisting of dataset watermarking and dataset verification. In dataset watermarking, some watermarked samples were generated and inserted into the dataset, as did in poisoning-based backdoor attacks. In dataset verification, a hypothesis test was conducted based on the predictions of benign and poisoned testing samples. 
The proposed protection method can not only verify whether an open-sourced dataset has been used for training a model, but also not influence the normal usage of the dataset. 
Extensive experiments have verified the effectiveness of the proposed method.

\newpage

\bibliographystyle{unsrt}
\bibliography{reference}

\begin{thebibliography}{10}

\bibitem{Zhu2020TheRO}
Hao Zhu, Yonatan Bisk, and G.~Neubig.
\newblock The return of lexical dependencies: Neural lexicalized pcfgs.
\newblock {\em Transactions of the Association for Computational Linguistics},
  2020.

\bibitem{zhang2019hibert}
Xingxing Zhang, Furu Wei, and Ming Zhou.
\newblock Hibert: Document level pre-training of hierarchical bidirectional
  transformers for document summarization.
\newblock In {\em ACL}, 2019.

\bibitem{ren2019fastspeech}
Yi~Ren, Yangjun Ruan, Xu~Tan, Tao Qin, Sheng Zhao, Zhou Zhao, and Tie-Yan Liu.
\newblock Fastspeech: Fast, robust and controllable text to speech.
\newblock In {\em NeurIPS}, 2019.

\bibitem{liu2020transferring}
Songxiang Liu, Yuewen Cao, Shiyin Kang, Na~Hu, Xunying Liu, Dan Su, Dong Yu,
  and Helen Meng.
\newblock Transferring source style in non-parallel voice conversion.
\newblock In {\em INTERSPEECH}, 2020.

\bibitem{C_H@TPAMI_2020}
C.~{Huang}, Y.~{Li}, C.~C. {Loy}, and X.~{Tang}.
\newblock Deep imbalanced learning for face recognition and attribute
  prediction.
\newblock {\em IEEE Transactions on Pattern Analysis and Machine Intelligence},
  42(11):2781--2794, 2020.

\bibitem{Wu_2019_ICCV}
Zuxuan Wu, Xin Wang, Joseph~E. Gonzalez, Tom Goldstein, and Larry~S. Davis.
\newblock Ace: Adapting to changing environments for semantic segmentation.
\newblock In {\em ICCV}, 2019.

\bibitem{deng2009imagenet}
Jia Deng, Wei Dong, Richard Socher, Li-Jia Li, Kai Li, and Li~Fei-Fei.
\newblock Imagenet: A large-scale hierarchical image database.
\newblock In {\em CVPR}, 2009.

\bibitem{cifar}
Alex Krizhevsky.
\newblock {\em Learning multiple layers of features from tiny images}, 2009.

\bibitem{k-anon@2002}
Latanya Sweeney.
\newblock k-anonymity: A model for protecting privacy.
\newblock {\em International Journal of Uncertainty, Fuzziness and
  Knowledge-Based Systems}, 10(05):557--570, 2002.

\bibitem{ZK@2019}
Z.~{Tu}, K.~{Zhao}, F.~{Xu}, Y.~{Li}, L.~{Su}, and D.~{Jin}.
\newblock Protecting trajectory from semantic attack considering
  ${k}$-anonymity, ${l}$-diversity, and ${t}$-closeness.
\newblock {\em IEEE Transactions on Network and Service Management},
  16(1):264--278, 2019.

\bibitem{YS@2019}
Y.~{Sei}, H.~{Okumura}, T.~{Takenouchi}, and A.~{Ohsuga}.
\newblock Anonymization of sensitive quasi-identifiers for l-diversity and
  t-closeness.
\newblock {\em IEEE Transactions on Dependable and Secure Computing},
  16(4):580--593, 2019.

\bibitem{HUA2019403}
Zhongyun Hua, Yicong Zhou, and Hejiao Huang.
\newblock Cosine-transform-based chaotic system for image encryption.
\newblock {\em Information Sciences}, 480:403 -- 419, 2019.

\bibitem{CHEN2019420}
Lanxiang Chen, Wai-Kong Lee, Chin-Chen Chang, Kim-Kwang~Raymond Choo, and Nan
  Zhang.
\newblock Blockchain based searchable encryption for electronic health record
  sharing.
\newblock {\em Future Generation Computer Systems}, 95:420 -- 429, 2019.

\bibitem{LI2019113}
Jiguo Li, Qihong Yu, and Yichen Zhang.
\newblock Hierarchical attribute based encryption with continuous
  leakage-resilience.
\newblock {\em Information Sciences}, 484:113 -- 134, 2019.

\bibitem{arsalan2017protection}
Muhammad Arsalan, Aqsa~Saeed Qureshi, Asifullah Khan, and Muttukrishnan
  Rajarajan.
\newblock Protection of medical images and patient related information in
  healthcare: Using an intelligent and reversible watermarking technique.
\newblock {\em Applied Soft Computing}, 51:168--179, 2017.

\bibitem{chai_2019}
H.~{Chai}, S.~{Yang}, Z.~L. {Jiang}, and X.~{Wang}.
\newblock A robust and reversible watermarking technique for relational dataset
  based on clustering.
\newblock In {\em TrustCom/BigDataSE}, pages 411--418, 2019.

\bibitem{D_Hu_2019}
D.~{Hu}, D.~{Zhao}, and S.~{Zheng}.
\newblock A new robust approach for reversible database watermarking with
  distortion control.
\newblock {\em IEEE Transactions on Knowledge and Data Engineering},
  31(6):1024--1037, 2019.

\bibitem{chen2017targeted}
Xinyun Chen, Chang Liu, Bo~Li, Kimberly Lu, and Dawn Song.
\newblock Targeted backdoor attacks on deep learning systems using data
  poisoning.
\newblock {\em arXiv preprint arXiv:1712.05526}, 2017.

\bibitem{gu2019badnets}
Tianyu Gu, Kang Liu, Brendan Dolan-Gavitt, and Siddharth Garg.
\newblock Badnets: Evaluating backdooring attacks on deep neural networks.
\newblock {\em IEEE Access}, 7:47230--47244, 2019.

\bibitem{bagdasaryan2020backdoor}
Eugene Bagdasaryan, Andreas Veit, Yiqing Hua, Deborah Estrin, and Vitaly
  Shmatikov.
\newblock How to backdoor federated learning.
\newblock In {\em AISTATS}, 2020.

\bibitem{li2020rethinking}
Yiming Li, Tongqing Zhai, Baoyuan Wu, Yong Jiang, Zhifeng Li, and Shutao Xia.
\newblock Rethinking the trigger of backdoor attack.
\newblock {\em arXiv preprint arXiv:2004.04692}, 2020.

\bibitem{liu2020reflection}
Yunfei Liu, Xingjun Ma, James Bailey, and Feng Lu.
\newblock Reflection backdoor: A natural backdoor attack on deep neural
  networks.
\newblock In {\em ECCV}, 2020.

\bibitem{li2020invisible}
Shaofeng Li, Minhui Xue, Benjamin Zhao, Haojin Zhu, and Xinpeng Zhang.
\newblock Invisible backdoor attacks on deep neural networks via steganography
  and regularization.
\newblock {\em IEEE Transactions on Dependable and Secure Computing}, 2020.

\bibitem{liu2020survey}
Yuntao Liu, Ankit Mondal, Abhishek Chakraborty, Michael Zuzak, Nina Jacobsen,
  Daniel Xing, and Ankur Srivastava.
\newblock A survey on neural trojans.
\newblock In {\em ISQED}, 2020.

\bibitem{li2020backdoor}
Yiming Li, Baoyuan Wu, Yong Jiang, Zhifeng Li, and Shu-Tao Xia.
\newblock Backdoor learning: A survey.
\newblock {\em arXiv preprint arXiv:2007.08745}, 2020.

\bibitem{hogg2005introduction}
Robert~V Hogg, Joseph McKean, and Allen~T Craig.
\newblock {\em Introduction to mathematical statistics}.
\newblock Pearson Education, 2005.

\bibitem{wilcoxon1992individual}
Frank Wilcoxon.
\newblock Individual comparisons by ranking methods.
\newblock In {\em Breakthroughs in statistics}, pages 196--202. Springer, 1992.

\bibitem{simonyan2014very}
Karen Simonyan and Andrew Zisserman.
\newblock Very deep convolutional networks for large-scale image recognition.
\newblock In {\em ICLR}, 2015.

\bibitem{he2016deep}
Kaiming He, Xiangyu Zhang, Shaoqing Ren, and Jian Sun.
\newblock Deep residual learning for image recognition.
\newblock In {\em CVPR}, 2016.

\bibitem{gtsrb}
Johannes Stallkamp, Marc Schlipsing, Jan Salmen, and Christian Igel.
\newblock Man vs. computer: Benchmarking machine learning algorithms for
  traffic sign recognition.
\newblock {\em Neural networks}, 32:323--332, 2012.

\bibitem{ilyas2019adversarial}
Andrew Ilyas, Shibani Santurkar, Dimitris Tsipras, Logan Engstrom, Brandon
  Tran, and Aleksander Madry.
\newblock Adversarial examples are not bugs, they are features.
\newblock In {\em NeurIPS}, 2019.

\end{thebibliography}

\end{document}